\documentclass{aa}

\usepackage[varg]{txfonts}
\usepackage{amsmath}
\usepackage{amssymb}
\usepackage{graphicx}
\usepackage{color}

\newcommand\porb{0.34008279}

\newcommand{\xmm}{\emph{XMM-Newton}}
\newcommand\ergscmsq{\,erg\,s$^{-1}$\,cm$^{-2}$}

\begin{document}

\title{V902 Monocerotis: a likely disc-accreting intermediate polar\thanks{Based on observations obtained at the Calar Alto observatory},\thanks{Based on observations obtained with XMM-Newton, an ESA science mission with instruments and contributions directly funded by ESA Member States and NASA}}

\author{
H. Worpel\inst{1}\fnmsep\thanks{Corresponding author:{hworpel@aip.de}},
A.~D. Schwope\inst{1},
I. Traulsen\inst{1},
K. Mukai\inst{2,3},
S. Ok\inst{4,1}
}
\institute{Leibniz-Institut f\"ur Astrophysik Potsdam (AIP), An der Sternwarte 16, 14482 Potsdam, Germany \and
           CRESST and X-ray Astrophysics Laboratory, NASA/GSFC, Greenbelt, MD 20771, USA\and
           Department of Physics, University of Maryland, Baltimore County, 1000 Hilltop Circle, Baltimore, MD 21250, USA\and
           Department of Astronomy \& Space Sciences, Faculty of Science, University of Ege, 35100, Bornova, \.Izmir, Turkey
}
\authorrunning{Worpel et al}
\date{}

\keywords{cataclysmic variables-- individual: V902 Monocerotis-- techniques: spectroscopic-- binaries: eclipsing-- X-rays: binaries  }

\abstract
{}
{We aim to confirm whether the eclipsing cataclysmic variable V902 Mon is an Intermediate Polar, to characterise its X-ray spectrum and flux, and to refine its orbital ephemeris and spin period.}
{We performed spectrographic observations of V902 Mon in 2016 with the 2.2\,m Calar Alto telescope, and X-ray photometry and spectroscopy with \emph{XMM-Newton} in October 2017. This data was supplemented by several years of AAVSO visual photometry.}
{We have confirmed V902 Mon as an IP based on detecting the spin period, with a value of 2,208\,s, at multiple epochs. Spectroscopy of the donor star and Gaia parallax yield a distance of $3.5^{+1.3}_{-0.9}$\, kpc, suggesting an X-ray luminosity one or two orders of magnitude lower than the $10^{33}$\,erg\,s$^{-1}$ typical of previously known IPs. The X-ray to optical flux ratio is also very low.

The inclination of the system is more than $79^\circ$, with a most likely value of around $82^\circ$. We have refined the eclipse ephemeris, stable over 14,000 cycles.
The H$\alpha$ line is present throughout the orbital cycle and is clearly present during eclipse, suggesting an origin distant from the white dwarf, and shows radial velocity variations at the orbital period. The amplitude and overall recessional velocity seem
inconsistent with an origin in the disc. The \emph{XMM-Newton} observation reveals a partially absorbed plasma model typical of magnetic CVs, with a fluorescent iron line at 6.4\,keV showing a large equivalent width of 1.4\,keV. 
}
{V902 Mon is an IP, and probably a member of the hypothesized X-ray underluminous class of IPs. It is likely to be a disc accretor, though the radial velocity behaviour of the H$\alpha$ line remains puzzling. The large equivalent width of the fluorescent iron line, the small $F_X/F_\text{opt}$ ratio, and the only marginal detection of X-ray eclipses suggests that the X-ray emission arises from scattering.}

\maketitle

\section{Introduction}

Intermediate polars (IPs) are cataclysmic variable stars (CVs), consisting of a strongly magnetised ($\sim 10$\,MG) white dwarf primary and a Roche-lobe filling secondary. The magnetic field of the primary prevents the formation of an inner accretion disc, and channels the infalling gas to relatively small regions near the primary's magnetic poles. The accreting gas forms standing shocks at those sites, leading to the formation of accretion columns and the release of strong bremsstrahlung-like X-ray emission. 

In contrast to their more magnetised relatives, the polars, IPs do not have synchronised spin and orbital periods. The detection of two distinct periodicities (usually $>3$\,hr for the orbit, and a few tens of minutes for the spin) is therefore a crucial observational diagnostic. Both classes, however, show strong emission lines (the Balmer hydrogen lines, neutral and ionized helium), leading to numerous discoveries in spectroscopic surveys.

V902 Monocerotis is an eclipsing CV, and probable IP, discovered in the Isaac Newton
Telescope (INT) Photometric H$\alpha$ Survey of the northern Galactic plane (IPHAS) survey by its prominent H$\alpha$ emission \citep{WithamEtAl2007}. It is optically faint compared to previously known IPs, with a white-light magnitude of approximately 16.5, and X-ray faint relative to its optical brightness with a flux of $2.2\times 10^{-13}$ erg\,s$^{-1}$\,cm$^{-2}$ between 0.5 and 10\,keV as determined from a short \emph{Swift} observation in 2009 \citep{AungwerojwitEtAl2012}.

V902 Mon has an orbital period of about 8.162 hours. A white dwarf (WD) spin period of around 2,210\,s was inferred from optical photometry \citep{AungwerojwitEtAl2012}, implying that the system is an intermediate polar CV, but the spin modulation was not evident at every epoch so the classification as an intermediate polar (IP) has so far not been completely secure (see e.g. \citealt{Patterson1994} for identification criteria).

Confirming V902 Mon as a member of the IP class would be valuable for a number of reasons. Both its
orbital and WD spin periods are atypically long, it would be the eclipsing IP with the
second-longest known period, beside Nova Scorpii 1437 \citep{SharaEtAl2017, PotterBuckley2018} and the longest period eclipsing IP not associated with a known past nova. Furthermore, its magnetic field strength of $\sim 18$ MG, as inferred from its orbital and spin periods empirically \citep{AungwerojwitEtAl2012}, is uncommonly strong for an IP. The magnetic field might be measurable under positive circumstances through the detection of a cyclotron component in the optical/IR spectral regime.

Its X-ray flux is low \citep{AungwerojwitEtAl2012}, potentially making V902 Mon a member of a hypothesised class of X-ray underluminous IPs \citep{PretoriusMukai2014} that may contribute much to the Galactic Ridge X-ray Emission (GRXE), a diffuse-appearing X-ray emission concentrated around the Galactic plane. This emission was originally speculated to be truly diffuse, but a \emph{Chandra} observation determined that around 80\% of it can be resolved into numerous individually faint point sources \citep{RevnivtsevEtAl2009}. Their nature is still uncertain but IPs may be the dominant contribution (e.g. \citealt{WarwickEtAl2014}), thus representing a crucial component of the X-ray emission of our Galaxy, and of other galaxies. IPs may, in fact, be the most common X-ray source class above a luminosity of $10^{31}$\,erg\,s$^{-1}$ \citep{PretoriusMukai2014}. Unlike many of the X-ray dim IP candidates, which are low in flux at all wavelengths due to (presumably) low accretion rate, V902 Mon appears to be X-ray weak compared to its optical brightness; see for example Figure 9 of \cite{AungwerojwitEtAl2012}.

We present a spectroscopic observation of V902 Mon obtained with the 2.2\,m telescope at the
Calar Alto observatory to attempt to verify the WD rotation period, to seek cyclotron
spectral features that allow the magnetic field strength to be measured, and to obtain spectroscopy of the donor star during eclipse for obtaining the system distance. 
We have also analysed an X-ray observation by XMM-Newton to seek spin modulations at X-ray wavelengths and to characterise the high energy spectrum of the system. Using
observations by the American Association of Variable Star Observers (AAVSO) we have refined the eclipse ephemeris and also verified the white dwarf rotation period over eight years. 

\section{Observations}

\subsection{Optical observations}

V902~Mon was observed for six hours on the night of Jan~9~2016, with the 2.2\,m telescope at the
Calar Alto observatory. We used the Calar Alto Faint Object Spectrograph (CAFOS) instrument with the g200 grism, and a slit width of 1.5". This instrument provided wavelength coverage of approximately 3750 \AA\ to
10500 \AA\ at a FWHM resolution of about 13 \AA , as measured from arc lamp spectra, though the usable wavelength range was narrower (about 5000 \AA\ to 9000 \AA) for fainter stellar spectra. The spectrograph was rotated so that the light of another star, a 14th magnitude star with designation PSO J062747.364+014843.893, was also on the slit. A total of 36 exposures were taken. The exposure length was 600\,s and we obtained continuous exposure, except for a $\sim$20\,min period where the telescope dome had to be closed due to strong wind. The eighth exposure in the sequence was cut short at 518\,s due to the closing of the dome. The readout time of the exposures was 56\,s. We reduced these spectra using the ESO-MIDAS software \citep{Warmels1992}. To correct for possible time-dependent flexure of the spectrograph we used the 5577.3387\,\AA\ atmospheric oxygen line.

While performing the data reduction we became aware of another faint star, PSO~J062746.70$+$014814.78, also lying on the slit. It has magnitudes $m_G=20.3, m_R=19.5, m_I=18.9, m_Z=18.7, m_Y=18$ in the Panoramic Survey Telescope and Rapid Response System (Pan-STARRS) survey \citep{ChambersEtAl2016}. It is redder than both V902 Mon and the comparison star and may give a small amount of stray light, since it is only 3" from V902 Mon, but this star is so faint that it is unlikely to give any significant contamination except possibly during eclipse.

According to the AAVSO Photometric All-Sky Survey-6 (APASS; \citealt{HendenEtAl2016}), the comparison star has a Johnson V magnitude of $13.932\pm0.016$. We used this information to obtain Calar Alto synthetic V-band photometry for V902 Mon by integrating the flux between 5,000 \AA\ and 5,900 \AA\ for both stars, under the assumption that both objects were properly centered on the slit at all times, such that the same photometric corrections apply to both V902 Mon and the comparison star.

We then downloaded 4,659 individual photometric observations of V902 Mon taken by AAVSO observers between 2008 March 24 and 2017 Nov 08. Most were in the $V$ or $CV$ bands, but some $I$-band observations were also performed in 2017 Oct. The timings were corrected to the Solar System barycenter using the algorithm of \cite{EastmanEtAl2010}. The AAVSO observations occurred mainly during five distinct observing epochs in 2008 Nov-2009 Jan, 2014 Feb-Mar, 2015 Dec, 2016 Jan-Mar, and 2017 Oct. The last of these epochs was performed specifically to support the XMM-Newton observation; see AAVSO Alert Notice \#601.

We also performed a 3 hour observation of V902 Mon with the 1\,m T100 telescope at TUBITAK National Observatory on the night of 2017 Sep 27-28. The observation was performed in V band, beginning with 80\,s exposures which was later reduced to 60\,s when the viewing conditions and airmass became more favourable. The data were reduced with IRAF \citep{Tody1993}.

\subsection{XMM-Newton observations}

V902 Mon was observed by \emph{XMM-Newton} on 2017 Oct 14 (Obs ID 0804111001). The EPIC cameras \citep{StruderEtAl2001, TurnerEtAl2001} were operated in full frame mode with the medium filter and the Optical Monitor (OM; \citealt{MasonEtAl2001}) observed in imaging and fast modes with the UVM2 filter. The Reflection Grating Spectrometer (RGS; \citealt{denHerderEtAl2001}) was operated in HER + SES modes.

Using the XMM-Newton Science Analysis Software (SAS), we reduced the EPIC-$pn$ and EPIC-MOS data with the \texttt{emchain} and \texttt{epchain} tasks respectively. The OM fast and imaging mode data were reduced with \texttt{omichain} and \texttt{omfchain}, and the RGS data were reduced with \texttt{rgsproc}, though due to the source's faintness we have not used the RGS data.

The duration of the observation was nominally 43\,ks and we obtained usable EPIC-$pn$ data from phase 0.03 in orbital cycle 13822 to phase 0.40 in cycle 13823. The observation was affected by a moderate amount of soft proton flaring. For the source extraction regions we used circles centered on the source with radii 15, 9.5, and 16 arcseconds for EPIC-$pn$, EPIC-MOS1, and EPIC-MOS2 respectively. The comparatively small extraction radius of MOS1 is because the source is directly adjacent to a row of dead pixels, and the total flux of the source is contained within a smaller area.

\section{Method and results}

\subsection{Optical/UV photometry}

\begin{figure}
 \includegraphics{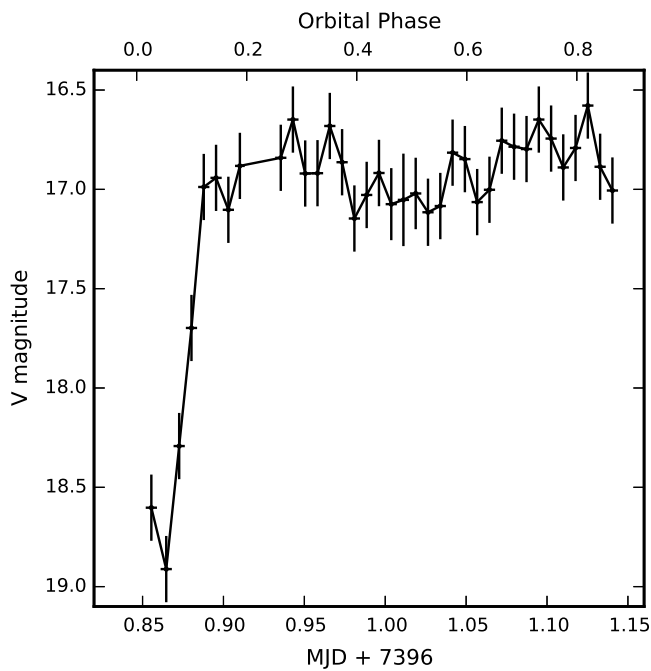}
 \caption{$m_V$ light curve of V902 Mon derived from Calar Alto spectroscopy}
 \label{fig:licu}
\end{figure}

In Figure \ref{fig:licu} we show the synthetic V-band photometry of V902 Mon taken at Calar Alto on 2016 Jan 9. The observations began during eclipse, which is two magnitudes fainter than the baseline brightness. There are indications of smaller oscillations on a shorter timescale than the orbital period.

We used the analysis-of-variance method \citep{Schwarzenberg-Czerny1989} to search for periodicities
in the AAVSO data. The orbital period of 8.162 hours was obvious, and there was a distinct peak around 2,210 seconds, confirming the previous result of \cite{AungwerojwitEtAl2012}. We refined the spin period measurement as follows. We removed the eclipses ($-0.08<\phi_\text{orb}<0.08$), and divided the data points into a subset for each of the four observational epochs. The Calar Alto points were included with those of the 2016 observing period. We calculated a spin period for each set, with uncertainties obtained using bootstrapping \citep{Efron1979} on a minimum of 1,000 trials. 

We used 25 AoV bins and a step size of $5\times 10^{-4}$\,s, corresponding to an accumulated error of approximately 1\,s over the length of the 2016 observational epoch (i.e., the longest one). For the five observational epochs we obtained best candidate spin periods of 2208.9$\pm$ 3.8\,s, 2208.6$\pm$1.1\,s, 2214.4$\pm$6.5\,s, 2208.3 $\pm$ 1.5\,s, and 2208.9 $\pm 5.1$\,s. Combining the results with a simple constant fit gives an overall spin period of 2208.62$\pm$0.85\,s. There was no evidence for the beat periods of $(1/P_\text{spin}-1/P_\text{orb})^{-1}=2,390$\,s or $(1/2P_\text{spin}-1/P_\text{orb})^{-1}=5,200$\,s, suggesting that V902 Mon accretes predominantly from a disc. 

The possibility that the 2,208\,s periodicity is an orbital sideband cannot be completely ruled out, though we did not find any periodicity at the implied "true" spin period of 2,054\,s. We think this possibility is unlikely, since not many IPs are known where the sidebands dominate completely, but even if so it would not affect the hypothesis of a spin period different from the orbital period.

\begin{figure}
 \includegraphics{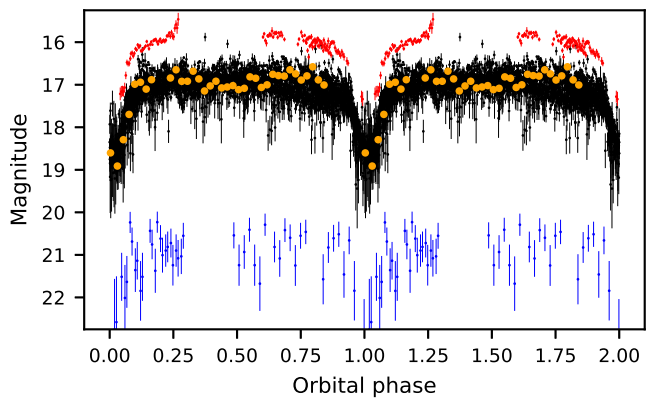}
 \caption{Light curves of V902 Mon folded on the refined orbital period of \porb\ days. Two cycles are plotted for clarity. Small black and red points are AAVSO V-band and I-band observations respectively, the large orange points are from the Calar Alto observation, and the blue points are XMM-Newton Optical Monitor data.}
 \label{fig:orbit_phasefold}
\end{figure}

The eclipses are clearly visible in AAVSO V and R band observations (see Figure \ref{fig:orbit_phasefold}). The XMM-Newton UVM2 magnitudes, obtained from 600\,s binning, appear in the same figure to also show the eclipse but it is less conspicuous due to the large uncertainties.

\begin{figure}
 \includegraphics{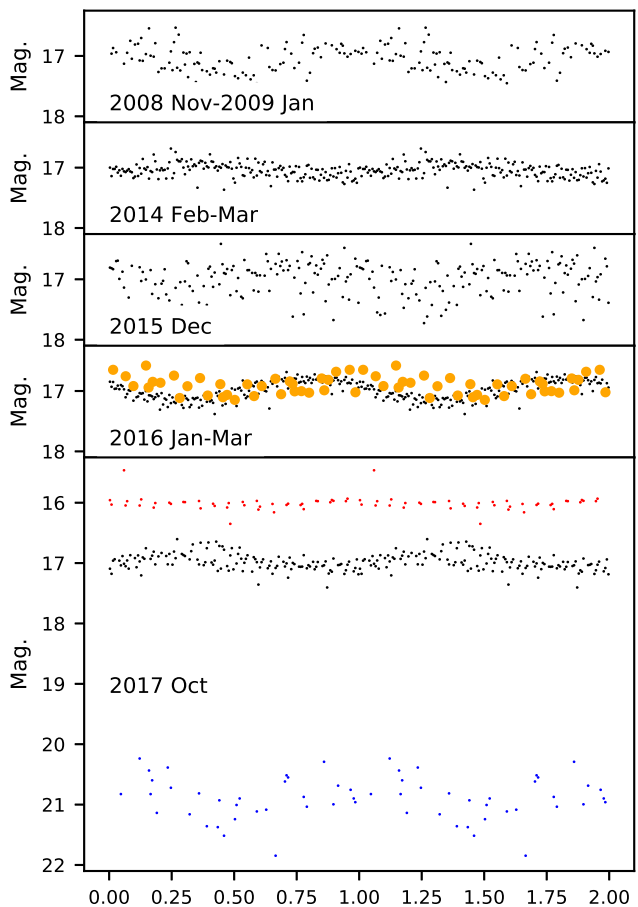}
 \caption{Light curves of V902 Mon folded on the inferred spin period of 2208\,s, with a binning of 200 bins per spin period. Two cycles are plotted for
          clarity and the eclipses have been excluded. Shown are the folded light curves for four epochs:
          2008 Nov 25 -- 2009 Jan 5 \emph{(top panel)},
          2014 Feb 18 -- 2014 Mar 16 \emph{(second panel)},
          2015 Dec 5  -- 2015 Dec 29 \emph{(third panel)},
          2016 Jan 9  -- 2016 Mar 16 \emph{(fourth panel)}, and
          2017 Oct 12 -- 2017 Oct 21 \emph{(fifth panel)}.
          A sinusoidal modulation is visible at all five epochs.  Large black and red points indicate the AAVSO V-band and I-band observations for the epoch in question. The Calar Alto data points are indicated with large orange markers in the fourth panel, and the XMM-Newton Optical Monitor data are the blue points in the bottom panel. The spin phase zero point is arbitrary.}
 \label{fig:spin_phasefold}
\end{figure}

As shown in Figure \ref{fig:spin_phasefold} the modulation in optical photometry is visible in all epochs. It was not evident, however, on every night in 2014. There is an indication that the Calar Alto data points are offset from the AAVSO points, by approximately 0.2 phase units. However, the uncertainty of 0.85\,s in the spin period corresponds to an accumulated error of about one cycle at the time of the Calar Alto observation, so we cannot be sure that this offset is real. The XMM-Newton UVM2 data also appear to show spin oscillations, but this conclusion is uncertain since the error bars are comparatively large. There is no evidence at all for spin oscillations in the AAVSO R band.

\begin{table}
\caption{Eclipse midpoints for V902 Mon}
   \begin{tabular}{lrl}
\hline
   Eclipse midpoint& Cycle & Reference\\
(BJD) & & \\
\hline
 2453340.50763 & 0 & \cite{WithamEtAl2007}\\
 2453342.54805 & 6 & \cite{WithamEtAl2007}\\
 2454092.42921 & 2211 & \cite{AungwerojwitEtAl2012}\\
 2454093.44951 & 2214 & \cite{AungwerojwitEtAl2012}\\
 2454387.62159 & 3079 & \cite{AungwerojwitEtAl2012}\\
 2454388.64167 & 3082 & \cite{AungwerojwitEtAl2012}\\
 2454389.66144 & 3085 & \cite{AungwerojwitEtAl2012}\\
 2455158.92916 & 5347 & \cite{AungwerojwitEtAl2012}\\
 2456707.664 & 9901 & This work (AAVSO)\\
 2456710.727 & 9910 & This work (AAVSO)\\
 2456717.528 & 9930 & This work (AAVSO)\\
 2456718.550 & 9933 & This work (AAVSO)\\
 2456719.568 & 9936 & This work (AAVSO)\\
 2456720.591 & 9939 & This work (AAVSO)\\
 2456721.610 & 9942 & This work (AAVSO)\\
 2457368.785 & 11845 & This work (AAVSO)\\
 2457385.792 & 11895 & This work (AAVSO)\\
 2457397.356 & 11929 & This work (Calar Alto)\\
 2457435.784 & 12042 & This work (AAVSO)\\
 2457436.804 & 12045 & This work (AAVSO)\\
 2457444.625 & 12068 & This work (AAVSO)\\
 2457444.969 & 12069 & This work (AAVSO)\\
 2457445.985 & 12072 & This work (AAVSO)\\
 2457447.006 & 12075 & This work (AAVSO)\\
 2457447.686 & 12077 & This work (AAVSO)\\
 2458044.874 & 13833 & This work (AAVSO)\\
 2458045.213 & 13834 & This work (AAVSO)\\
   \end{tabular}
\label{tab:eclipse_times}
\end{table}

\begin{figure}
    \includegraphics{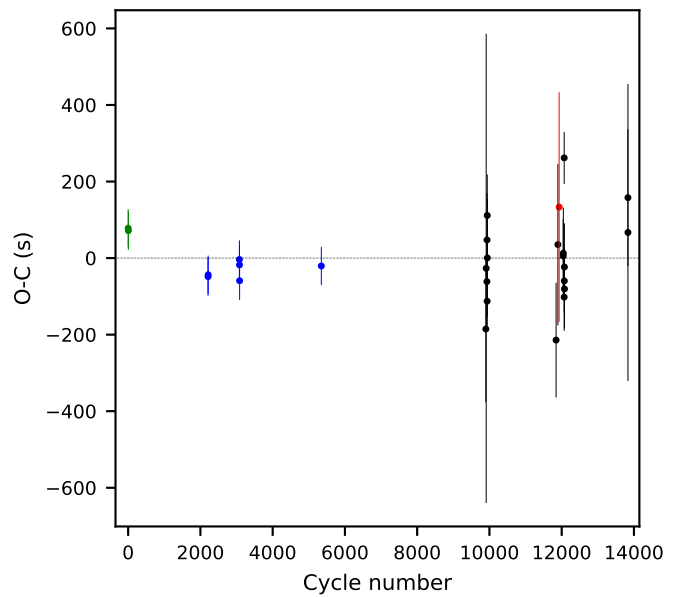}
    \caption{Observed minus calculated residuals to the eclipse timing fit (Equation \ref{eqn:ephem}). The green, blue, black, and red points are from \cite{WithamEtAl2007}, \cite{AungwerojwitEtAl2012}, AAVSO, and Calar Alto respectively. }
    \label{fig:ominusc}
\end{figure}

One eclipse is evident in the Calar Alto photometry, and twenty in the AAVSO data. Of these, eighteen are sufficiently well observed to allow eclipse timing. We calculated eclipse midpoints by fitting the measured magnitudes around the times of the eclipses with a constant plus a Gaussian. We then combined the timings of the eclipse minima from our data with the previous timings of \cite{WithamEtAl2007} and \cite{AungwerojwitEtAl2012} as summarised in Table \ref{tab:eclipse_times}. We fit the resulting timings with a linear ephemeris. For the previously reported data we assumed an uncertainty of 50\,seconds. We conservatively set an uncertainty of five minutes for the Calar Alto data point, and use either the formal uncertainties arising from the fits for the AAVSO data or 180\,s, whichever is greater. The updated ephemeris thus becomes 
 \begin{equation}
 T_0 = \text{BJD } 2453340.5067(3) + \porb(4)\times E
 \label{eqn:ephem}
\end{equation}
with $\chi^2_\nu=1.3$ for the goodness of fit. Residuals to the fit are shown in Figure \ref{fig:ominusc}. Aside from an apparent systematic offset for the timings reported in \cite{WithamEtAl2007}, no obvious trend is visible, and the observations lack the precision to search for residuals caused by, e.g., a circumbinary planet. The uncertainty obtained for the orbital period is much less than one cycle over the 13,834 cycles reported in Table \ref{tab:eclipse_times}.

 Many of the eclipse profiles in \cite{AungwerojwitEtAl2012} support the hypothesis that the accretion disc is often still partially visible; those with longer durations resemble the smooth eclipse profiles in UX~UMa stars, e.g., \cite{Smak1994}, where the disc contributes a substantial fraction of the total system luminosity. The shorter, flat-bottomed eclipses more resemble the covering of a point-like emitter, perhaps suggesting that at those times the accretion disc contains less mass and contributes a smaller proportion of the total system luminosity. A similar variable disc extent has been inferred for the IP FO~Aqr by \cite{HameuryLasota2017}.
 
Interestingly, when the eclipses in \cite{AungwerojwitEtAl2012} have flat bases, the spin modulations seem to be more subdued than at times when the eclipse profiles are smoother. This phenomenon may have implications for the accretion geometry, as we discuss later in section \ref{sec:conclusion}. Unfortunately we cannot conclusively verify it in the AAVSO data because during eclipses the system is often too faint for small telescopes to easily observe, particularly the very faint flat-based ones.
 
\begin{figure}
 \includegraphics{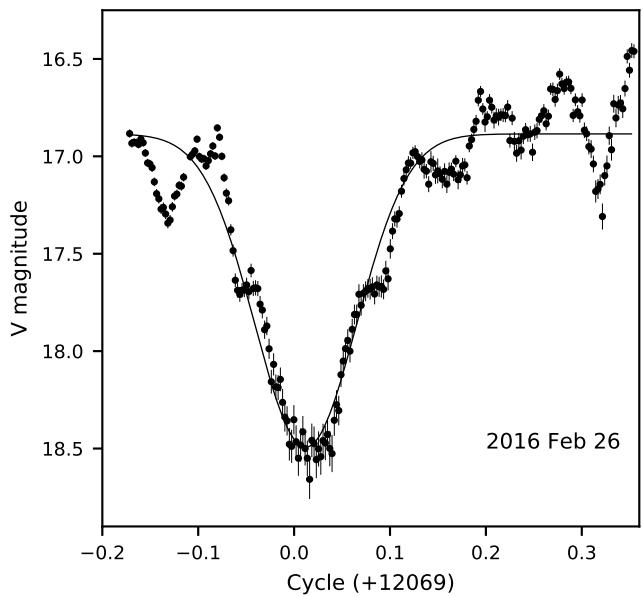}
 \caption{AAVSO light curve of one eclipse, observed in 2016 Feb, showing clear spin oscillations during eclipse ingress and egress. Also shown (black curve) is the Gaussian fit used for the eclipse timings}
 \label{fig:eclosc}
\end{figure}

We can use the existence of eclipses and the Roche lobe geometry of the system to estimate its inclination. From visual inspection of the phase-folded AAVSO data we confirm the estimate by \cite{AungwerojwitEtAl2012} that the full width of the eclipse at half-depth $0.106< \Delta\phi_{1/2}<0.120$. The major uncertainty in this estimate is that the baseline level is hard to determine with spin period modulations superimposed on it. If we assume that the shorter, flat-bottomed, eclipses occur when there is little contribution by the disc we can treat the accretion region on the star as a point source. One round-based eclipse profile is shown in Figure \ref{fig:eclosc}. This eclipse shows pronounced rotational oscillations, including during eclipse ingress and egress.

 \cite{ChananEtAl1976} give formulae for calculating the inclination given $\Delta\phi_{1/2}$ and the mass ratio $q=M_2/M_1$ for the eclipse of a point-like source by a Roche lobe shaped obscuring object. Since stable mass flow from secondary to primary requires $q<1$, we find that $79^\circ<i<87^\circ$ for the shorter eclipse durations and highest secondary mass. Assuming that the mass of the secondary is $0.65M_\odot<M_2<0.87M_\odot$, and $M_2<M_1<1.06M_\odot$ (\cite{Ramsay2000} determined the mean WD mass in an IP system to be $0.85\pm0.21$), the most likely value for the inclination is $81.7^\circ\pm1.6^\circ$. This estimate is similar to the previously reported estimate in \cite{AungwerojwitEtAl2012}; they inferred $77^\circ < i < 84^\circ$. This measurement, of course, depends on the assumption that the white dwarf is completely eclipsed. We argue that total obscuration of the white dwarf is very likely despite the ambiguous X-ray evidence (see Sect.~\ref{sec:xraylicu}), on the basis of the flat eclipse profiles during low accretion states and the deep ($\sim 1.5$\,mag) eclipses that obscure around 75\% of the optical flux.

\begin{figure}
 \includegraphics{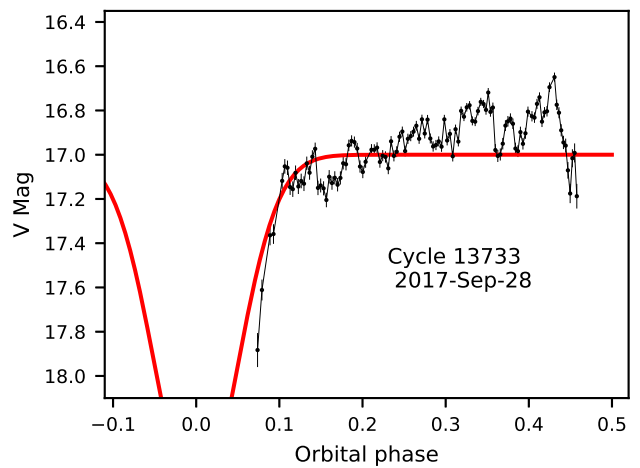}
 \caption{$V$-band light curve of the Sep 28 TUBITAK observation, with the eclipse according to our updated ephemeris indicated as the red curve.}
 \label{fig:tubitak}
\end{figure}

Our Calar Alto observations do not have the timing resolution to determine whether the eclipse we observed had a flat or round base. We believe that it occurred during the flat-bottomed profile regime for several reasons. The magnitude near eclipse midpoint was around $m_V=19.0$ rather than the $m_V=18.5$ typical of the smoother eclipse profile. Secondly, the spin modulations were smaller in amplitude than at the smooth profiled eclipse times in \cite{AungwerojwitEtAl2012}. As shown in Figure \ref{fig:tubitak}, the short observation at TUBITAK National Observatory suggests that the period of small spin oscillations lasted at least several weeks, from late September to late October 2017. That observation also shows the steep and deep eclipse egress profile associated with this accretion state.

\subsection{Optical spectroscopy}
\label{sec:spectroscopy}

\begin{figure}
 \includegraphics{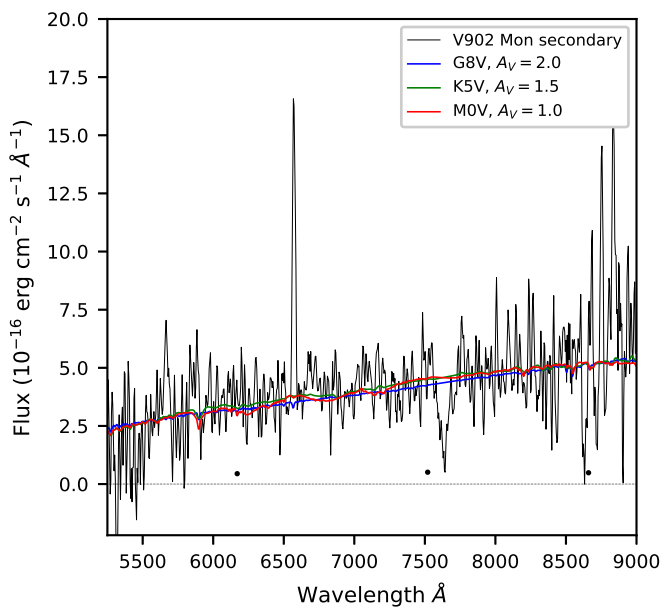}
 \caption{ Mean eclipse spectrum (black curve) of V902 Mon, and a range of stellar spectra from \cite{Pickles1998} for several values of visual extinction $A_V$. The telluric lines around 7,600 \AA\ have not been corrected in the V902 Mon spectrum, and the H$\alpha$ line is very prominent. The large black points represent the maximum possible contamination due to the nearby faint star, based on its Pan-STARRS magnitudes in the r,i,z filters, and this contribution is included in the stellar spectra shown.}
 \label{fig:eclipse_spectrum}
\end{figure}

During eclipse the primary is obscured by the secondary, as is the irradiated hemisphere of the secondary. The eclipses therefore provide the means to observe the companion without the white dwarf, and with a minimum of disc emission. We extracted the three eclipse spectra and averaged them to get a representative eclipse spectrum. Then we compared it to stellar spectra from \cite{Pickles1998}, corrected for interstellar reddening according to the formulae of \cite{CardelliEtAl1989}. For this exercise we adopted an $R_V$ parameter of 3.1, the value in the diffuse interstellar medium, though the results in the optical/near infrared are almost independent of the value chosen; see \cite{CardelliEtAl1989} for details. According to \cite{AungwerojwitEtAl2012}, the best-fitting $N_H$ is approximately $5\times 10^{21}$\,cm$^{-2}$, which implies a visual extinction of 2.8 magnitudes \citep{PredehlSchmitt1995}. However, such a large visual extinction can only be accommodated at the red end of the spectrum with a companion star brighter than G0V, i.e., more massive than the primary. Such a configuration is highly unlikely to give stable mass transfer, but lower values of the extinction give decent fits for a G8V spectrum at $A_V=2.0$ to an M0V red dwarf with $A_V=1.0$; see Fig \ref{fig:eclipse_spectrum}. This result gives a companion mass in the range 0.65-0.87$M_\odot$ (see Table 3 in \citealt{Knigge2006}), with intermediate values most likely since we probably have substantial interstellar absorption as well as some remaining contribution from the disc.

Assuming the G8V-M0V identification of the secondary is correct, we have magnitudes of $M_V=5.6-8.8, m_V\sim 18.5-19.0$ for the secondary and we thereby calculated a distance to the source of approximately $870-4,800$\,pc. This makes V902 Mon a potentially distant IP candidate. We note, however, that an unknown amount of stray light from the nearby red star means that V902 Mon's secondary is probably towards the bluer end of this range, making the lowest distances somewhat less likely.

The Gaia Data Release 2 \citep{Gaia2016, Gaia2018} gives a parallax for V902 Mon of $0.285\pm 0.125$\,mas. Interpreted naively, this would
imply a distance of $3.5^{+2.7}_{-1.1}$\,kpc. However, as pointed out by \cite{Bailer-Jones2015}, such estimates become unreliable when the
proportional error in the parallax is greater than about 20\%. More sensible results are obtained if a normalisable prior on the
distance is used. Fortunately, our knowledge of the spectrum of the secondary star provides us with such a prior. We will take a top-hat distribution with these limits as the prior. The
refined distance estimate then becomes $3.5^{+1.3}_{-0.9}$\,kpc. The catalogue of \cite{Bailer-JonesEtAl2018}, which uses a distance prior based on an exponentially decreasing space density prior, gives a very similar distance of $3.0^{+1.6}_{-0.9}$\,kpc. Furthermore, using this prior implicitly assumes that CVs are drawn from the same space density distribution as the general stellar population. CVs are an older population and should therefore have a larger scale height, and so slightly larger distances are favoured.

\begin{figure}
 \includegraphics{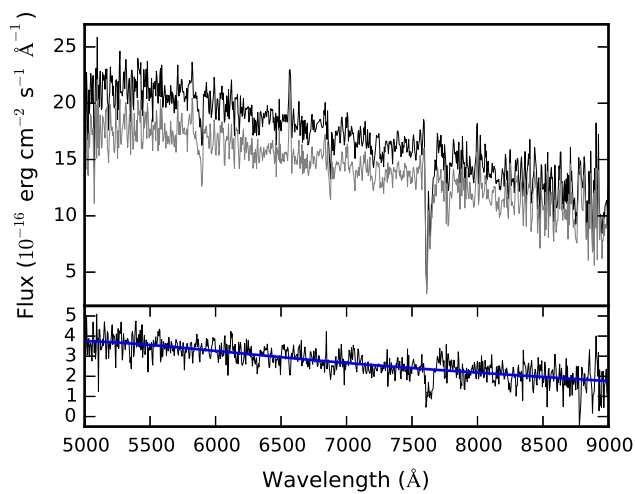}
 \caption{Bright and faint spin phase spectra (top panel; black and grey curves) and their difference (bottom panel). Both have had the mean eclipse spectrum (Fig \ref{fig:eclipse_spectrum})
          subtracted. The spectra have not been corrected for telluric absorption.  A blackbody fit of 17,800\,K ($A_V=2.0$) is shown in the bottom panel (blue curve).}
 \label{fig:spin_spectrum}
\end{figure}

To investigate the behaviour of the spectra with respect to the spin period, we divided the non-eclipse spectra into two groups according to their $\phi_\text{spin}$.
Although the epoch for the spin ephemeris is arbitrary, we found that $0.5<\phi_\text{spin}<1.0$ conveniently divides the Calar Alto data points into bright and
faint groups (see Figure \ref{fig:spin_phasefold}). We took mean spectra for both groups and subtracted the mean eclipse spectrum from them. The results are shown
in Figure \ref{fig:spin_spectrum}. It is clear from the difference spectrum that the brighter points are also bluer, exactly as would be expected from an accreting hot spot.

We fitted the difference spectrum with a reddened blackbody and found a best-fit temperature of 6,100$\pm 70$\,K for $A_V=0$ and $17,800\pm 800$\,K for $A_V=2.0$. Both alternatives fit the data equally well. 

There is no obvious evidence for cyclotron humps in the spectra, and we are therefore unable to directly measure the strength of the magnetic field at the poles. If the emission there were dominated by the cyclotron spectrum produced by a 15-20\,MG magnetic field, we would expect to see a red spectrum rather than the blue one actually detected. We therefore conclude that the magnetic field strength of V902 Mon is less than 15\,MG, consistent with the hypothesis that it is an IP.

\subsection{H$\alpha$ emission line properties}

\begin{figure}
 \includegraphics{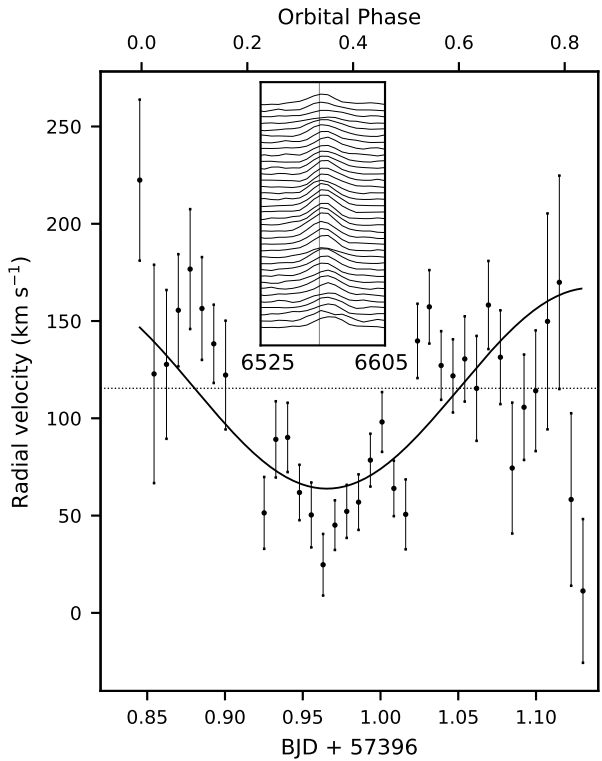}
 \caption{Radial velocity as measured from the H$\alpha$ line, with the best-fitting sinusoid shown. The inset shows the vicinity (6525-6605\AA) of the H$\alpha$ line in chronological order from top to bottom, with the $6562.8$\AA\ rest wavelength of H$\alpha$ indicated with a vertical line.}
 \label{fig:redshift}
\end{figure}

We used the H$\alpha$ line to calculate radial velocities in the V902 Mon system. Each spectrum was fit between 6,000 and 9,000 \AA, excluding ranges of telluric absorption, with a linear continuum plus a Gaussian near 6563 \AA\ representing the H$\alpha$ line to obtain central wavelengths and associated uncertainties for that emission line. {Though often multiple Gaussians are present, out relatively poor FWHM of 13\AA\ make trying to distinguish them a hopeless task; see the inset to Figure \ref{fig:redshift} and e.g. \cite{HellierEtAl1990} who required 1.2\AA\ to observe multiple features for FO~Aqr.}

We then converted the resulting wavelengths to radial velocities, and corrected for Earth's motion using the algorithms presented in \cite{WrightEastman2014}. The velocities were then fitted with a sinusoid plus constant to get both the orbital velocities and overall line-of-sight motion of the V902 Mon system. This fit is shown in Figure \ref{fig:redshift}. To the eye it appears as though there may be a shorter period, lower amplitude oscillation in addition to the orbital modulation. However, these features are irregularly aligned in phase and separated by considerably more than the spin period, so we conclude that they are just noise.

We found an overall recessional velocity of $115\pm 16$\,km\,s$^{-1}$, and an orbital modulation of $v\sin i=52\pm 18$\,km\,s$^{-1}$ with an offset of $0.60\pm0.11$ in phase. The amplitude of the orbital modulation is not consistent with an origin centered on the WD; for stellar masses of $M_D=0.8M_\odot, M_C=0.65M_\odot$ we would expect a much higher amplitude of $\sim 150$\,km\,s$^{-1}$. 

\begin{figure}
 \includegraphics{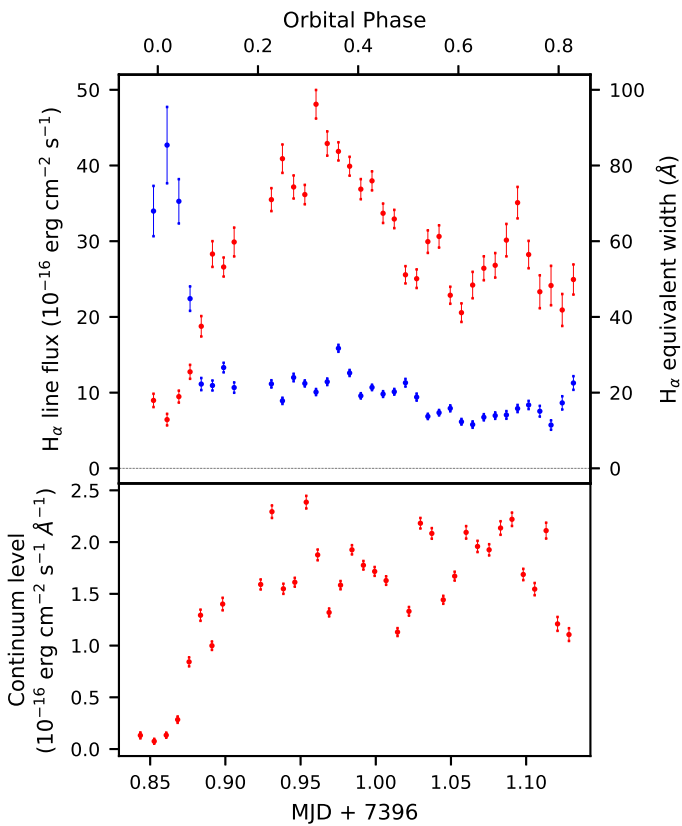}
 \caption{Flux (red) and equivalent width (blue) of the H$\alpha$ line.}
 \label{fig:hafluxeqw}
\end{figure}

The behaviour of the H$\alpha$ line flux and equivalent width are shown in Figure \ref{fig:hafluxeqw}. The line flux decreases during eclipse but does not go to zero, consistent with our previous findings, but its equivalent width increases strongly during eclipse. This is further evidence that parts of the H$\alpha$ line originate from sites rather distant from the WD. There is no other obvious orbital phase dependence, except possibly a slight decrease around phases 0.6-0.8, for the equivalent width. There is no spin period dependence, suggesting no relationship with the accreting spots on the WD surface, and completely consistent with its absence from the bottom panel of Figure \ref{fig:spin_spectrum}.

\subsection{X-ray photometry}
\label{sec:xraylicu}

V902 Mon was faint in the XMM-Newton X-ray data. By visual inspection of the light curves we found no evidence of variability, even during the eclipse. We used the Rayleigh method (e.g. \citealt{Brazier1994}) to search for a periodicity in the photon arrival times near the likely 2,208\,s spin period but no signal was detected.

To determine if there was any evidence of the eclipse at all we split the photon lists into eclipse ($-0.05<\phi_\text{orb}<0.05$) and non-eclipse sections according to the ephemeris of equation \ref{eqn:ephem}. For this exercise the extraction radii were 125, 110, and 150\,px (6.25", 5.5", and 7.5")  for EPIC-$pn$, MOS1, and MOS2 respectively, smaller than those for the spectral extractions to ensure minimal contamination by the soft proton flaring. We found evidence for a lower X-ray flux during eclipse in EPIC-$pn$ and MOS1, but not with MOS2. 

We also sought spin period modulations in a similar manner: if the apparent UV spin modulation seen in Figure \ref{fig:spin_phasefold} is real then the X-ray count rate from spin phase 0.75 to 1.25 should exceed the count rate from orbital phase 0.25 to 0.75, which we indeed found for EPIC-$pn$ and EPIC-MOS1 but not MOS2. The results are summarised in Table \ref{tab:photcountrates}. We therefore have only tentative evidence for the eclipse or spin modulations in X-rays. It may be that much of the X-ray emission originates far from the white dwarf, as was found for the prototypical X-ray faint IP DQ Herculis \citep{MukaiEtAl2003}.

\begin{table}
 \caption{Photon count rates, in photons per kilosecond, for the eclipse and out-of-eclipse orbital phases, and for two spin period phase selections.}
 \begin{tabular}{l|lll}
  & PN & MOS1 & MOS2 \\
  \hline
  $0.05<\phi_\text{orb} <0.95$ & $9.6\pm0.7$&  $2.5\pm0.3$ & $3.3\pm0.4$\\
  $0.95<\phi_\text{orb} <1.05$ & $4.5\pm2.4$&  $0.4\pm0.8$ & $3.5\pm1.5$\\
  $0.75<\phi_\text{spin}<1.25$ & $13.0\pm1.1$& $2.9\pm0.5$ & $3.5\pm0.5$\\
  $0.25<\phi_\text{spin}<0.75$ & $7.9\pm0.9$& $2.1\pm0.4$  & $3.1\pm0.5$\\
 \end{tabular}

 \label{tab:photcountrates}
\end{table}

We also re-analyzed the X-ray data from the 2009 \emph{Swift} observation of V902 Mon originally presented in \cite{AungwerojwitEtAl2012}. The data were reduced with the \emph{xrtpipeline} tool. We used an 11 pixel (26") circle centered on the source as the extraction region, and a large circle in a nearby source-free area as the background extraction region. Only 24 photons in total were detected from V902 Mon. We then extracted a list of photon arrival times, corrected to the Solar System barycenter using the \emph{barycorr} task in HEASOFT, excluded the eclipses to avoid these interfering with the measurement of the spin period, and performed a period search using the Rayleigh method (e.g., \citealt{Brazier1994}). With so few photons it is unlikely that we could find a spin period modulation and, indeed, the only significant periodicities we detected were the 95.9 minute \emph{Swift} orbital period and its harmonics.

The Swift Ultraviolet/Optical Telescope \citep{RomingEtAl2005} was operated in imaging mode during the observation. We can therefore obtain no useful timing data from it, though a UV magnitude of 21 could be determined (see Fig \ref{fig:sed}).

\subsection{X-ray spectroscopy}
\label{sec:xrayspec}

\begin{table}
  \caption{X-ray spectral fit. Fluxes are bolometric. Note that we determined the 6.4\,keV iron line equivalent width by the method described later in the text. No upper limit to the temperature of the warmer plasma component could be determined.}
 \begin{tabular}{ll}
  Parameter & Value\\
  \hline\\
  $n_\text{H}$ & $2.63^{+4.15}_{-1.66}\times 10^{22}$\,cm$^{-2}$\\[6pt]
  Covering Fraction & $0.88^{+0.12}_{-0.21}$ \\[6pt]
  kT$_1$ & $194^{+55}_{-85}$\,eV \\[6pt]
  norm$_1$ & $3.64^{+9.59}_{-2.64}\times 10^{-5}$ \\[6pt]
  kT$_2$ & $15.36_{-1.69}$\,keV \\[6pt]
  norm$_2$ & $7.17^{+6.27}_{-2.53}\times 10^{-5}$ \\[6pt]
  Line energy & $6.48^{+0.18}_{-0.11}$\,keV\\[6pt]
  Line $\sigma$ & $0.12^{+0.27}_{-0.12}$\,keV\\[6pt]
  Line norm & $3.84^{+2.27}_{-1.91}\times 10^{-6}$\\[6pt]
  Line eqw & $1.43$\,keV\\[6pt]
  \hline\\
  $\chi^2_\nu$ & 0.90 (67 dofs) \\[6pt]
  Absorbed flux & $1.93\pm0.27 \times 10^{-13}$\, \ergscmsq\\[2pt]
  Unabsorbed flux & $4.43\pm0.61 \times 10^{-13}$\,\ergscmsq\\[2pt]
  
 \end{tabular}

 \label{tab:specfit}

\end{table}

\begin{figure}
 \includegraphics{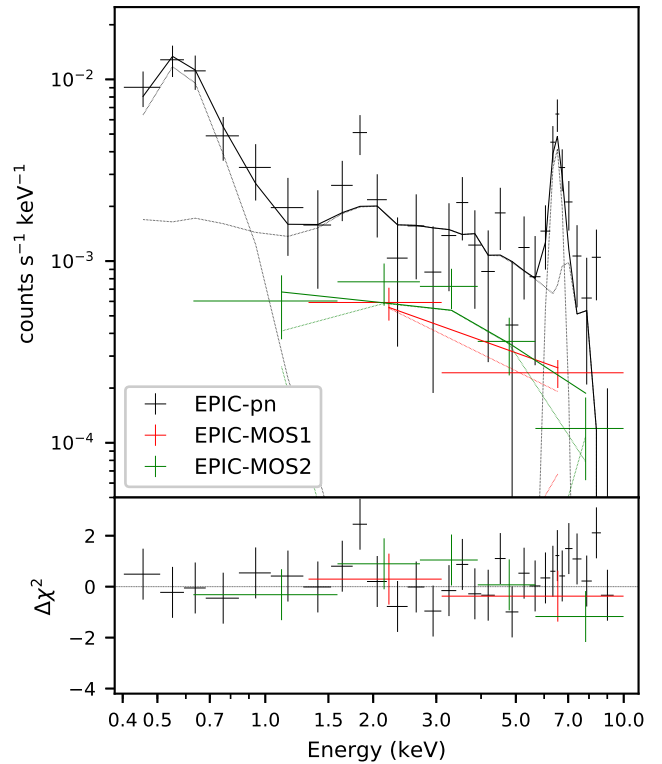}
 \caption{X-ray spectrum (\emph{top panel}) and $\chi^2$ residuals to the fit (\emph{bottom panel})}
 \label{fig:xray_spectrum}
\end{figure}

We extracted phase-averaged spectra for all three EPIC instruments and binned them with a minimum of 25 counts per bin. We used Xspec version 12.9.1q \cite{Arnaud1996} to fit the resulting spectra jointly, between approximately 0.4 and 10\,keV, with a partially covered two-temperature plasma atmosphere (i.e., two Mekals; \citealt{MeweEtAl1985,LiedahlEtAl1995}) plus a Gaussian around 6.4\,keV. The lower fit limit arises from the coarse spectral binning and low counts at soft energies. The results are shown in Figure \ref{fig:xray_spectrum}; though for clarity we have plotted the spectra with 50 counts per bin. Errors on the parameters were obtained with the \emph{steppar} command and are reported at the 99\% confidence level, and fluxes are bolometric. The results of the spectral fit are shown in Table \ref{tab:specfit}.

We also obtained an absorbed flux of $1.4\times 10^{-13}$\, \ergscmsq\ in the 0.5-10\,keV band, approximately two thirds of the brightness ($2.2\times 10^{-13}$\,erg\,s$^{-1}$\,cm$^{-2}$ in the $0.5-10$\,keV energy band) measured by \cite{AungwerojwitEtAl2012}. We have also obtained a somewhat lower value for the $n_H$; they obtained $4\times 10^{22}$\,cm$^{-2}$. These facts suggest that V902 Mon's disc contained less material than in 2012, leading to a lower accretion rate and reduced absorption.

Our distance estimate of $3.5^{+1.3}_{-0.9}$\,kpc gives an X-ray luminosity of approximately $(1.8 - 6) \times 10^{32}$\,erg\,s$^{-1}$ for the earlier Swift observation. This result establishes that V902 Mon is X-ray underluminous, by over one order of magnitude (e.g., \citealt{PretoriusMukai2014}), and moderately fainter still in the XMM-Newton observation.

To determine whether V902 Mon is X-ray faint relative to its optical flux, as suggested in \cite{AungwerojwitEtAl2012}, or whether it is faint overall, we constructed a crude X-ray to optical ratio by approximating the optical flux by $\log_{10}(F_\text{opt})=-m_\text{V}/2.5-5.37$ \citep{MaccacaroEtAl1988} and calculating the X-ray flux between 0.5 and 2.0\,keV. We obtained $\log_{10}(F_\text{X}/F_\text{opt})=-1.7$. For IPs generally, this ratio is around $+0.15$ on average (Schwope \& Thomas 2018, private communication), and for the low accretion rate IPs EX~Hya, YY~Dra, V1025~Cen, and HT~Cam we consistently got values of around +0.5. V902 Mon is thus X-ray faint relative to its optical flux by some two orders of magnitude.

\begin{figure}
 \includegraphics{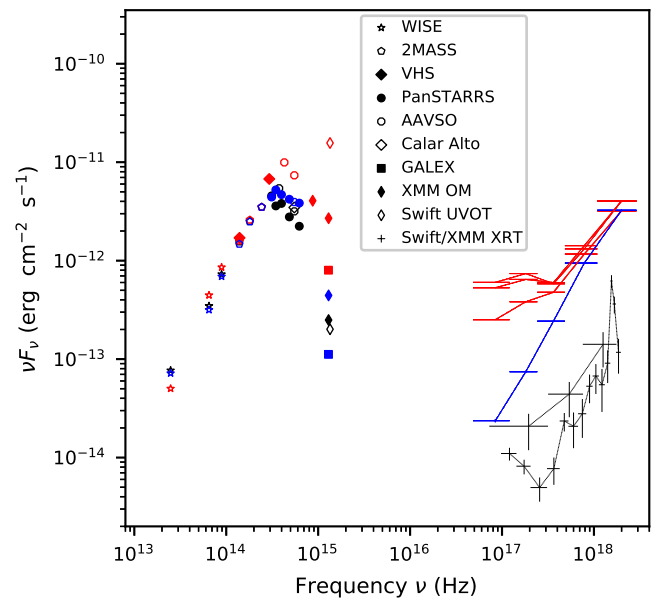}
 \caption{Spectral energy distribution of V902 Mon from infrared to X-rays (black), compared to the SEDs of V2069~Cyg and NY Lup (blue and red respectively). They have been shifted vertically to coincide with V902 Mon at the 2MASS J point.}
 \label{fig:sed}
\end{figure}

In Figure \ref{fig:sed} we show a spectral energy distribution for V902 Mon, from the infrared to X-rays. 
We also plotted the equivalent SEDs for the confirmed IPs V2069~Cyg and NY Lup. They have similar orbital periods to V902 Mon (7.48 and 9.87 hours respectively: \citealt{ThorstensenTaylor2001, deMartinoEtAl2006}) and can therefore be expected to have similar primary and donor stars. Their SEDs have been scaled to coincide with V902 Mon in the 2MASS J band (see below). The scaling factor for V2069~Cyg agrees, within measurement uncertainties, with the distance correction for its Gaia distance of 1,180\,pc. The scaling factor for NY~Lup (1,270\,pc) is however too large by a factor of four, suggesting that its secondary is fainter than those of V902~Mon and V2069~Cyg. Nonetheless, the shape of its SED is similar to the other two stars and, since it is $F_\text{X}/F_\text{opt}$ we ultimately are interested in, we will continue with the assumed scaling. The X-ray points for V2069~Cyg and NY~Lup are taken from the 3XMM-DR7 serendipitous source catalogue \citep{RosenEtAl2016}.

To construct the SEDs we used additional archival data from the Wide-field Infrared Survey Explorer (WISE), the 2-Micron All-Sky Survey (2MASS), the VISTA Hemisphere Survey (VHS), the Pan-STARRS survey, and GALEX, using the effective wavelengths and zero points for each filter given in \cite{JarrettEtAl2011}, \cite{CohenEtAl2003}, the Cambridge University Survey Unit database\footnote{\url{http://casu.ast.cam.ac.uk/surveys-projects/vista/technical/filter-set}},\cite{TonryEtAl2012}, and \cite{MorrisseyEtAl2007} respectively. The V-band Calar Alto data point is the mean of our non-eclipse spectra assuming an effective wavelength of 5450\,\AA. For the \emph{Swift} UVOT M2 point we use the AB magnitude listed in the catalogue of \cite{PageEtAl2014}, using the effective wavelength given in \cite{PooleEtAl2008}. The three \emph{Swift} X-ray points contain eight photons apiece, and their flux densities were obtained by considering each bin individually and fitting them with a constant. V902 Mon was not visible in the Burst Alert Telescope \citep{BarthelmyEtAl2005} image, so we have not included this data here. The effective wavelength and zero point for the \xmm\ OM filters are given in \cite{KirschEtAl2004}. We have used a coarser binning for the XMM-Newton points than we used in Figure \ref{fig:xray_spectrum}, to avoid cluttering the graph.

V902~Mon is approximately $1.5$ orders of magnitude fainter than the other two IPs at frequencies above $3\times 10^{17}$\,Hz. NY Lup furthermore shows a similar excess in soft X-rays to V902 Mon, with variable intensity, whereas V2069~Cyg appears to lack this feature. In the infrared the SEDs of all three sources appear very similar, with source-to-source and epoch-to-epoch variability becoming apparent in the optical to ultraviolet. We have thus determined that V902 Mon is genuinely X-ray underluminous compared to both IPs that are faint due to low accretion rate, and compared to IPs with similar orbital configurations.

Figure \ref{fig:xray_spectrum} shows a strong iron line at 6.4\,keV. We determined is equivalent width by fitting the EPIC$-pn$, MOS1, and MOS2 data jointly between 5.5 and 7.5\,keV with a bremsstrahlung spectrum plus three Gaussians representing the three iron lines. We used unbinned spectra to avoid losing fine spectral features and utilised the $c$-statistic for fitting. The temperature of the bremsstrahlung was held fixed at 15\,keV, and the energies of the helium-like and hydrogen-like iron lines were fixed to 6.698 and 6.927\,keV respectively (cf. Table 1 of \citealt{MeweEtAl1985}). The energy of the fluorescence line was allowed to float around 6.4\,keV. The widths of the lines were constrained to all be equal, but this value was not fixed. All four components had variable normalisation. The fit was satisfactory, with a $c$-stat of 341.5; the method of \cite{Kaastra2017} predicts 312.6$\pm 18.6$ for the same model. We found an equivalent with for the fluorescent iron line of 1.43\,keV, much wider than the 150\,eV one typically finds for magnetic CVs (eg \citealt{Mukai2017}), and more indicative of an X-ray emission dominated by scattering. The scattering interpretation also explains why the X-ray eclipses are only marginally detectable: the scattering surface, presumably the disc, is never fully obscured and leaving the possibility that the X-ray source is never seen directly.

\section{Conclusion}
\label{sec:conclusion}
We have found, using AAVSO and Calar Alto data, that the 2,208 second periodicity attributed to the WD spin was present at multiple epochs between 2008 and 2016. We have also found slight, but inconclusive, indications that the spin modulation is present in X-rays. This result effectively secures V902 Mon's status as an intermediate polar. Our observations have refined both the orbital ephemeris and the spin period.

There is strong evidence that V902 Mon possesses a disc. Many of its eclipses show a symmetric profile consistent with the covering of an extended emitter (see Figure 2 of \citealt{AungwerojwitEtAl2012}), and H$\alpha$ emission was still present during eclipse, suggesting an origin in the disc. The high probability that the X-ray emission is scattered also requires a physically large scattering medium, such as a disc. V902 Mon appears to accrete almost entirely from the disc, since in optical there is no evidence for the beat period that would represent direct coupling of gas to the magnetic field lines. 

However, the behaviour of the radial velocities of the H$\alpha$ line is puzzling: the amplitude of the modulations, though possessing the correct orbital period, is too small to represent the orbital speed of the WD and is probably offset in phase. Higher resolution optical spectroscopy will be necessary to determine its origin, and we recommend choosing a different comparison star to avoid contamination by the faint nearby star.

The magnitude of the spin modulations in optical seems to depend on the eclipse profile, with pronounced oscillations occurring when the eclipse profile is rounded and less conspicuous oscillations occurring when the eclipse profile is deeper and flatter. This phenomenon probably depends on numerous factors including the magnetic field geometry and the (variable) radius of the inner edge of the disc.

We have attempted to identify the spectral type of the secondary, but found that a wide range of stellar masses are good fits for plausible values of the interstellar absorption. A secondary as bright as G8V or as faint as M0V are possible with an intermediate value, perhaps around K5V, being most likely. The estimated distance of 3,500\,pc implies that V902 Mon is X-ray underluminous by around one order of magnitude compared to its optical luminosity. We obtained a similar X-ray deficit by comparing V902 Mon's SED with those of presumably similar IPs. This is a distinct phenomenon to IPs that are faint at all wavelengths because they are accreting at a low rate.

A lower mass secondary apparently conflicts with the estimate of 0.87-0.97 solar masses obtained by \cite{AungwerojwitEtAl2012} from applying the empirical mass-period relation in \cite{SmithDhillon1998}, but there is considerable scatter at higher orbital period in the data points used to obtain that relation. The apparent discrepancy is therefore not likely to be a problem. If, however, both a faint spectral type and high mass are correct, it would imply that the secondary is an evolved star rather than a main sequence star. That would also be consistent with a low accretion rate.

The spin modulations in optical are more pronounced when the disc contains more material, yet the X-ray spectrum seems to show that only about half of the flux is lost because of absorption. From these lines of evidence, one might think that obscuration is not the reason for V902~Mon's low $F_X/F_\text{opt}$ ratio. However, the lack of obvious X-ray eclipses and the enormous equivalent width of the 6.4\,keV iron line, suggest that the X-ray emission is primarily scattered. That is, the direct component is likely hidden by the disc, and the intrinsic luminosity is higher than we observe.

We have determined the inclination of V902 Mon to be greater than $79^\circ$, assuming the WD is eclipsed, with a most probable value of around $82^\circ$, but no greater than $87^\circ$. This value is well consistent with previous estimates. Higher resolution spectroscopy of the companion star during eclipse, preferably at a time when the accretion disc is nearly empty, would allow a better determination of V902 Mon's distance than is currently available with Gaia DR2.

\begin{acknowledgements}
This work was supported by the German DLR under contracts 50 OR 1405 and 50 OR 1711. We acknowledge with thanks the variable and comparison star observations from the AAVSO International Database contributed
by observers worldwide and used in this research. This research has made use of the APASS database, located at the AAVSO web site. Funding for APASS has been provided by the Robert Martin Ayers Sciences Fund. This work has made use of Astropy \citep{Astropy2013, Astropy2018}. This work has made use of data from the European Space Agency (ESA)
mission {\it Gaia} (\url{https://www.cosmos.esa.int/gaia}), processed by
the {\it Gaia} Data Processing and Analysis Consortium (DPAC,
\url{https://www.cosmos.esa.int/web/gaia/dpac/consortium}). Funding
for the DPAC has been provided by national institutions, in particular
the institutions participating in the {\it Gaia} Multilateral Agreement.
Samet Ok is supported by TUBITAK 2214-A International Doctoral Research Fellowship Programme. We thank TUBITAK for a partial support in using the T100 telescope with project number 16BT100-1027. We are grateful to the anonymous referee, whose suggestions improved the clarity of the paper.

\end{acknowledgements}

\bibliographystyle{aa}
\bibliography{bibli}

\end{document}